\documentclass[aps,prb,showpacs,groupedaddress,twocolumn]{revtex4}

\usepackage{graphicx}
\begin{document}

\title{Quantum Andreev effect in 2D HgTe/CdTe quantum well-superconductor systems}

\author{Qing-feng Sun$^{1}$}
\email[Email address: ]{sunqf@aphy.iphy.ac.cn}
\author{Yu-Xian Li$^{2}$}
\author{Wen Long$^{3}$}
\author{Jian Wang$^{4}$}
\email[Email address: ]{jianwang@hkusua.hku.hk}

\affiliation{ $^1$Institute of Physics, Chinese
Academy of Sciences, Beijing 100190, China\\
$^2$College of Physics, Hebei Normal University, Shijiazhuang
050016, China\\
$^3$Department of Physics, Capital Normal University, Beijing
100048, China\\
$^4$Department of Physics and the center of theoretical and
computational physics, The University of Hong Kong, Hong Kong, China
 }

\date{\today}

\begin{abstract}
The Andreev reflection (AR) in 2D HgTe/CdTe quantum
well-superconductor hybrid systems is studied. A quantized AR with
AR coefficient equal to one is predicted, which is due to the
multi-Andreev reflection near the interface of the hybrid system.
Importantly, this quantized AR is not only universal, i.e.,
independent of any system parameters and quality of the coupling of
the hybrid system, it is also robust against disorder as well. As a
result of this quantum Andreev effect, the conductance exhibits a
quantized plateau when the external bias is less the superconductor
gap.
\end{abstract}


\maketitle

Recently, the topological insulator (TI), a new state of matter, has
attracted a lot of theoretical and experimental
attention.\cite{ref1,ref2,ref3,ref4,ref5,ref6,ref7,ref8} The TI has
an insulating energy gap in the bulk states, but it has exotic
gapless metallic states on its edges or surfaces. The TI is first
predicted in two-dimensional (2D) systems, e.g. the graphene and
HgTe/CdTe quantum well (QW).\cite{ref2,ref3} The 2D TI has the
gapless helical edge states and exhibits the quantum spin Hall
effect. This helical edge states, with the opposite spins on a given
edge or opposite edges for a given spin direction containing
opposite propagation directions, are topologically protected and are
robust against all time-reversal-invariant impurities. Soon after
that, people also found the TI in three-dimensional (3D) materials,
e.g., $Bi_{1-x}Sb_x$, $Bi_2 Se_3$, etc.\cite{ref4,ref5} On the
experimental side, the TI in HgTe/CdTe QW, $Bi_{1-x}Sb_x$, $Bi_2
Se_3$, and so on, have been successfully
confirmed.\cite{ref4,ref6,ref7}

The Andreev reflection (AR) which was found about fifty years ago is
an important transport process.\cite{ref9} The AR occurs near the
interface of a conductor and a superconductor, in which an incident
electron from the metallic side is reflected as a hole and a Cooper
pair is created in the superconductor. When the bias is smaller than
the superconductor gap, the conductance of the
conductor-superconductor hybrid device is mainly determined by the
AR.\cite{aref1} For a perfect conductor-superconductor interface,
the probability of the AR can reach one. However, the AR coefficient
is usually very small due to various scattering mechanisms such as
the contact potential of the interface, the impurities, the mismatch
of the density of states of the conductor and superconductor, and so
on. Although the AR has been extensively investigated in various
conductor-superconductor hybrid systems, up to now, people has not
observed a robust quantized AR that holds for various system
parameters and sustains in the presence of a variety of scattering
mechanisms.

In this letter, we study the AR in a 2D TI-superconductor (TI-S)
hybrid system. We found that in this system the AR is quantized with
AR coefficient being one, because of the multi-Andreev reflection
along the TI-S interface. Importantly, this quantized AR is
universal and robust. The quantization of AR persists regardless of
the system parameters such as the Fermi energy, energy of the
incident electron and the size of the system. It is also robust
against the presence of impurities and contact potential of the
interface. Due to the quantum Andreev effect, the conductance
exhibits the plateau with the plateau value being $4e^2/h$ ($2
e^2/h$) for the two-terminal (four-terminal) TI-S hybrid system.

Since the TI phase in HgTe/CdTe QW has been experimentally
realized,\cite{ref6,ref7} we shall focus on the HgTe/CdTe QW in the
following calculation.  The results will be the same for other 2D TI
systems. We consider two HgTe/CdTe QW-superconductor hybrid devices
as shown in Fig.1(f): one is a ribbon of HgTe/CdTe QW coupled to a
semi-infinite superconducting lead which is referred as the
two-terminal device and the other is a cross of HgTe/CdTe QW coupled
to a superconducting lead which is referred as the four-terminal
device. The advantage of the four-terminal device is that the
"trajectories" of the incident electron and reflected hole can be
clearly shown.

The hybrid devices are described by the Hamiltonian $H=H_{TI}
+H_S+H_C$, where $H_{TI}$, $H_S$, and $H_C$ are the Hamiltonian of
the HgTe/CdTe QW, superconducting lead, and the coupling between
them, respectively. By discretizing spatial coordinates of the
continuous effective Hamiltonian $H_{TI}$ in Ref.\onlinecite{ref3}
and using the Nambu representation, the Hamiltonian $H_{TI}$ is
given by:\cite{ref11}
\begin{equation}
H_{TI} =\sum\limits_{{\bf i}}  \Psi^{\dagger}_{\bf i}
 \breve{H}_{\bf ii} \Psi_{\bf i} +
 \sum\limits_{\vec{\alpha}=(\vec{\delta}x,\vec{\delta}y),{\bf i}} \Psi^{\dagger}_{\bf i}
 \breve{H}_{{\bf i}\vec{\alpha}} \Psi_{{\bf i}+\vec{\alpha}}
 +H.c. ,
\end{equation}
where ${\bf i}=(i_x, i_y)$ is the site index, $\vec{\delta} x$ and
$\vec{\delta}y$ are unit vectors along $x$ and $y$ directions,
$\Psi_{\bf i} =(a_{\bf i}, b_{\bf i}, c^{\dagger}_{\bf i},
d^{\dagger}_{\bf i})^T$, and $a_{\bf i}, b_{\bf i}, c_{\bf i},
d_{\bf i}$ are annihilation operators of electron on the site ${\bf
i}$ at the states $|s,\uparrow\rangle$, $|p_x+ip_y,\uparrow\rangle$,
$|s,\downarrow\rangle$, and $-|p_x-ip_y,\downarrow\rangle$. In
Eq.(1), $\breve{H}_{{\bf ii}/{\bf i}\vec{\delta}x/{\bf
i}\vec{\delta}y} =\left(\begin{array}{ll} H_{{\bf ii}/{\bf
i}\vec{\delta}x/{\bf i}\vec{\delta}y} & {\bf 0}\\ {\bf 0} & -H_{{\bf
ii}/{\bf i}\vec{\delta}x/{\bf i}\vec{\delta}y} \end{array}\right)$
are the $4\times 4$ matrix Hamiltonian, where $H_{\bf ii}
=\left(\begin{array}{ll} E_s & 0\\0 & E_p\end{array}\right)$,
$H_{{\bf i}\vec{\delta}x} =\left(\begin{array}{ll} t_{ss} &
-it_{sp}\\-it_{sp} & t_{pp}\end{array}\right)$, and $H_{{\bf
i}\vec{\delta}y} =\left(\begin{array}{ll} t_{ss} & -t_{sp}\\t_{sp} &
t_{pp}\end{array}\right)$, with $E_{s/p} = C\pm M -E_F -4(D\pm
B)/a^2$, $t_{ss/pp}=(D\pm B)/a^2$, $t_{sp}=A/2a$. Here $E_F$ is the
Fermi energy (pinned by superconductor condensate), $a$ is the
lattice constant, and $A$, $B$, $C$, $D$, and $M$ are the system's
parameters which can be experimentally controlled. The Hamiltonian
$H_S$ of the superconducting lead is:
$
 H_S =\Sigma_{{\bf k},\sigma} \epsilon_{\bf k} a^{\dagger}_{S{\bf k}\sigma} a_{S{\bf k}\sigma}
  +\Sigma_{\bf k} \Delta (a^{\dagger}_{S{\bf k} \uparrow}
  a^{\dagger}_{S-{\bf k}\downarrow} + H.c.
) $ where $\Delta$ is the superconductor gap and $a^{\dagger}_{S{\bf
k}\sigma}$ ($a_{S{\bf k}\sigma}$) is the creation (annihilation)
operators in the superconducting lead with the momentum ${\bf
k}=(k_x,k_y)$. Here we consider the general s-wave superconductor.
The coupling Hamiltonian $H_C$ is: $H_C = \sum_{\bf
i}(a^{\dagger}_{S i_x \uparrow}, a_{S i_x \downarrow}) {\bf t}_S
\Psi_{\bf i} +H.c. $, where the operator $a_{S i_x\sigma} =\sum_{\bf
k} e^{i k_x i_x a} a_{S{\bf k}\sigma}$ and $
{\bf t}_S = \left( \begin{array}{llll} t_{Sa} & t_{Sb} &0 &0\\
   0& 0& -t_{Sa} & -t_{Sb} \end{array}\right).
$ Here the parameters $t_{Sa}$ and $t_{Sb}$ are the coupling
strengths between the superconductor and HgTe/CdTe QW, which depends
on the interface's contact potential and the quality of the coupling
in the experiment.

By using the Green's functions, the charge current $I_{ne}$ and spin
current $I_{se}$ from the n-th terminal of the HgTe/CdTe QW flowing
into the device are $I_{ne} =e (I_{n\uparrow}+I_{n\downarrow})$ and
$I_{ns} =(\hbar/2) (I_{n\uparrow}-I_{n\downarrow})$,
where\cite{ref12}
\begin{eqnarray}
 I_{n\sigma} & =& \frac{1}{h} \int dE \left\{
  \sum\limits_m T_{nm\sigma} (f_{n\sigma}-f_{m\sigma}) +
  T_{ns\sigma}(f_{n\sigma} -f_s) \right. \nonumber\\
& & \left. +\sum\limits_m
  T^A_{n\sigma,m\bar{\sigma}} (f_{n\sigma}-f_{m\bar{\sigma}}) \right\}.
\end{eqnarray}
Here $\bar{\sigma}=\downarrow,\uparrow$ for
$\sigma=\uparrow,\downarrow$, $f_{n\uparrow/\downarrow}(E) =f(E\mp
eV_n)$, and $f_{s}(E) =f(E)$, with $f(E)$ being the Fermi
distribution function and $V_n$ being the voltage of the terminal-n.
In Eq.(2), $T_{nm\sigma}(E) = \mathrm{Tr} \{ {\bf \Gamma}_{n\sigma}
{\bf G}^r_{\sigma\sigma} {\bf \Gamma}_{m\sigma} {\bf
G}^a_{\sigma\sigma} \}$ and $T_{ns\sigma}(E) = \mathrm{Tr} \{ {\bf
\Gamma}_{n\sigma} [{\bf G}^r {\bf \Gamma}_{s} {\bf
G}^a]_{\sigma\sigma} \}$ are respectively the normal transmission
coefficient from the terminal-n to the terminal-m and to the
superconductor terminal, and $T^A_{n\sigma, m\bar{\sigma}}(E) =
\mathrm{Tr} \{ {\bf \Gamma}_{n\sigma} {\bf G}^r_{\sigma\bar{\sigma}}
{\bf \Gamma}_{m\bar{\sigma}} {\bf G}^a_{\bar{\sigma}\sigma} \}$ is
the AR coefficient with the incident electron from the terminal-n
and the reflected hole going to the terminal-m. The linewidth
function ${\bf \Gamma}_{n/s}(E) =i[{\bf \Sigma}^r_{n/s}
-({\bf\Sigma}^r_{n/s})^{\dagger}]$ and the Green's functions ${\bf
G}^{r/a}(E)$ can be calculated from ${\bf G}^r(E) =({\bf
G}^a(E))^{\dagger} =\{E-{\bf H}_{cen}
-\sum_n{\bf\Sigma}^r_n-{\bf\Sigma}^r_s\}^{-1}$, where ${\bf
H}_{cen}$ is the Hamiltonian of the scattering region  as shown in
Fig.1f (dotted region). The self-energy functions
${\bf\Sigma}^r_n(E)$ and ${\bf\Sigma}^r_s(E)$ due to terminals of
the TI and superconducting lead can be calculated as in
Ref.[\onlinecite{ref12,ref13,ref14}]. In the following numerical
calculations, we choose the parameters from the realistic
materials:\cite{ref6} (1) the HgTe/CdTe QW's parameters are $A=364.5
meVnm$, $B=-686meVnm^2$, $C=0$, and $D=-512 meV nm^2$. (2) the
superconductor's parameters are the gap energy $\Delta=1meV$. The
lattice constant $a$ is set to $5nm$ and the TI-S coupling strengths
are taken $t_{Sa}=t_{Sb}\equiv t$.

\begin{figure}
\includegraphics[width=8cm,totalheight=8cm]{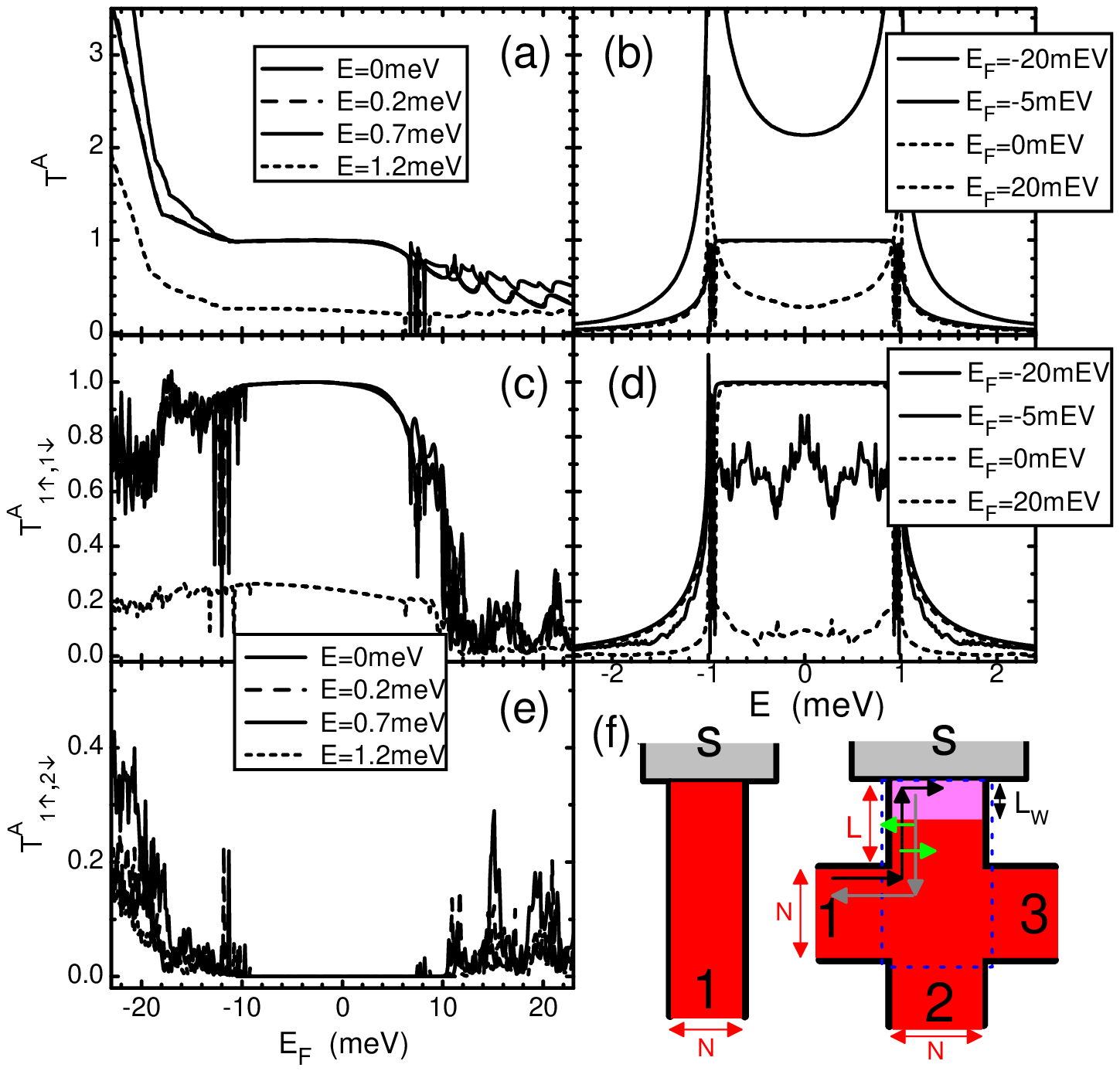}
\caption{ (Color online) (a) and (b): The AR coefficient $T^A$ vs.
$E_F$ (a) and energy $E$ (b) in the two-terminal device. (c), (d),
and (e): $T^A_{1\uparrow,1\downarrow}$ vs. $E_F$ (c),
$T^A_{1\uparrow,1\downarrow}$ vs. the energy $E$ (d), and
$T^A_{1\uparrow,2\downarrow}$ vs. $E_F$ (e) in the four-terminal
device with the length $L=750nm$. The other parameters in panels
(a)-(e) are: the width $N=300nm$, coupling strength $t=3meV$, and
$M=-10meV$. (f) is the schematic diagram for the two-terminal and
four-terminal devices. } \label{fig:1}
\end{figure}

We first study the two-terminal system. Due to the time-reversal
symmetry and the $C_2$ symmetry around the y axis, the AR
coefficients have the properties
$T^A_{1\uparrow,1\downarrow}(E)=T^A_{1\downarrow,1\uparrow}(E)\equiv
T^A(E)$ and $T^A(E)=T^A(-E)$. Fig.1a and 1b show the AR coefficient
$T^A$ versus the Fermi energy $E_F$ and energy $E$ of the incident
electron. Now the parameter $M$ is set $-10meV$ so that the
HgTe/CdTe QW is in the TI phase.\cite{ref3} Fig.1a and 1b clearly
show that $T^A$ exhibits a plateau with the plateau value equal to
one as long as $E_F$ is in the bulk gap of TI and the energy $E$
within the superconductor gap $\Delta$. Since there is only one
transmission channel (i.e. the edge state) when $E_F$ is in the bulk
gap, $T^A=1$ means that the incident electron is completely Andreev
reflected. It is remarkable that the quantum Andreev effect occurs
in such a wide range of $E_F$ and $E$, while in all previous
metal-superconductor hybrid systems the resonant condition $T^A=1$
occurs only at certain set of parameters. Fig.1a also shows that
when $E_F$ is out of the bulk gap, there is no plateau in the AR
coefficient $T^A$ and $T^A$ depends on both $E_F$ and $E$. For
$E_F<M$, $T^A$ can be larger than one (see Fig.1a) because there are
many transmission channels. However, the AR coefficient for each
transmission channel is still much smaller than one.

In order to reveal the nature of quantum Andreev effect, we will
study the four-terminal device. In the four-terminal device, the AR
coefficient $T^A_{n\uparrow,m\downarrow}$ has $3\times 3=9$
elements, so the trajectories of the incident electron and reflected
hole can clearly be shown. When the HgTe/CdTe QW is in the TI
regime, only $T^A_{1\uparrow,1\downarrow}$ is nonzero and other
eight elements are zero. In particular, as soon as the energy
$|E|<\Delta$ and $M<E_F<-M$, the quantized plateau with
$T^A_{1\uparrow,1\downarrow}=1$ emerges and this plateau is
independent of $E_F$ and $E$ (see Fig.1c and 1d). These results can
be understood with the help of helical edge states, in which the
spin-up and spin-down carriers move along the edge of the HgTe/CdTe
QW in clockwise and counter-clockwise directions,
respectively.\cite{ref1,ref3}  We consider the case of the spin-up
electron coming from the terminal-1 when the energy $|E|<\Delta$. As
shown in Fig.1f, two reflection processes occur at the TI-S
interface: (1). it can be Andreev reflected back as a spin-down hole
to the same terminal which will contribute to
$T^A_{1\uparrow,1\downarrow}$. (2). It can also be normal reflected
as an electron (spin up) along the TI-S interface, eventually to
terminal-3. Note that the normal reflection as an electron back to
the terminal-1 is prohibited by the time-reversal invariance and the
helical edge states being a pair of Kramer states. The reflected
electron traveling to terminal-3 has to go along the TI-S interface
since the only available transmission channel is the edge state.
This results in a reflection again at the TI-S interface where part
of the electron is Andreev reflected as the hole back to terminal-1
and the rest is normal reflected as electron towards terminal-3. As
this continues, multi-reflections occur as normal electron traverses
along the TI-S interface and eventually the transmission probability
$T_{13}$ becomes zero if the TI-S interface is long enough. Clearly,
it is this multi-AR that gives rise to the quantum Andreev effect
with the quantized AR $T^A_{1\uparrow,1\downarrow}=1$. Furthermore,
we have three observations: (i) If the energy $|E|>\Delta$, all AR
coefficients decrease as usual (see Fig.1d),\cite{ref15} because of
the occurrence of the normal tunneling from TI to the
superconductor. (ii) When $E_F$ is out of the bulk gap, all AR
coefficients have the same behavior: AR coefficient for each
transmission channel is in general small and strongly depends on
$E_F$ and $E$ (see Fig.1c, 1d, and 1e). (iii) For the spin-down
incident carrier, it is easy to show that
$T^A_{3\downarrow,3\uparrow}$ has the quantized plateau similar to
$T^A_{1\uparrow,1\downarrow}$.

\begin{figure}
\includegraphics[width=8.5cm,totalheight=4cm]{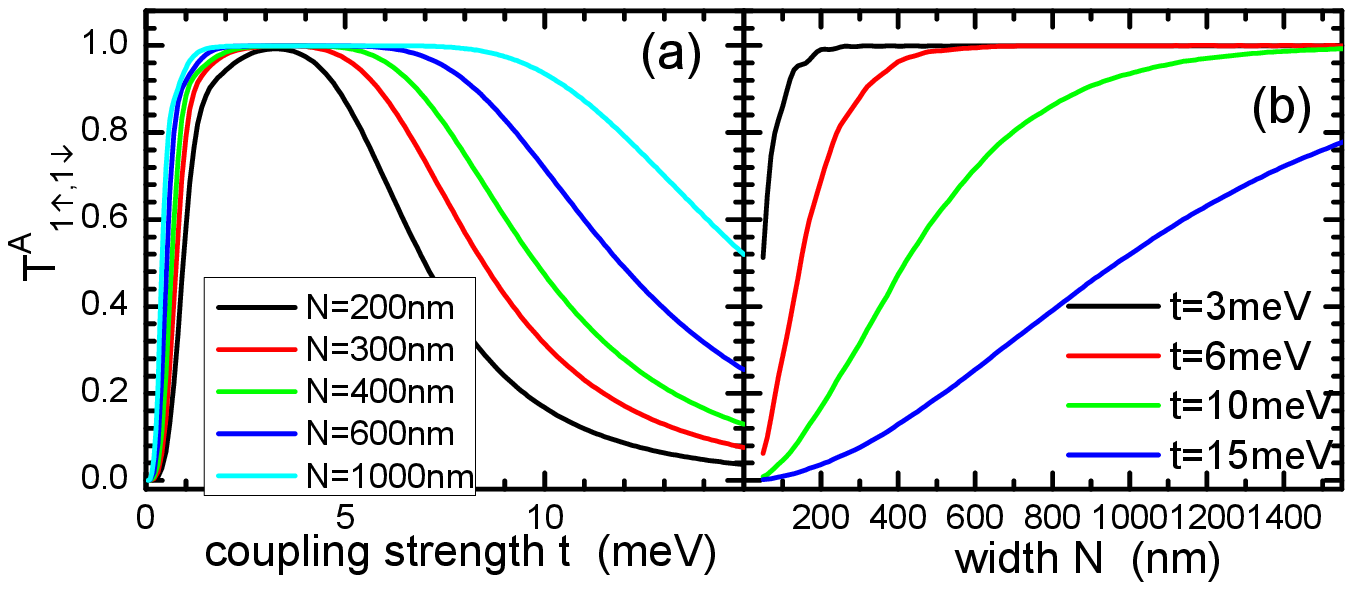}
\caption{ (Color online) $T^A_{1\uparrow,1\downarrow}$ vs. $t$ (a)
and width $N$ (b) in the four-terminal device with $E_F=-5meV$,
$E=0.1meV$, $M=-10meV$, and length $L=750nm$. } \label{fig:2}
\end{figure}

Next, we study how the quantized AR is affected by the system
parameters. Fig.2 shows the AR coefficient
$T^A_{1\uparrow,1\downarrow}$ versus the width $N$ of the HgTe/CdTe
QW ribbon and the coupling strength $t$. The results show that the
quantization of AR persists for a broad range of the coupling
strength $t$. In addition, the wider the width $N$, the broader the
quantization plateau is. For $N=1000nm$, the AR quantization plateau
can sustain when $t$ varies nearly one order of magnitude. Finally,
with the increase of the width $N$, the AR coefficient
$T^A_{1\uparrow,1\downarrow}$ rises monotonously before it reaches
quantized value due to the fact that the incident electron has more
chance of multi-AR for the longer TI-S interface. These results show
the universal feature of the quantum Andreev effect: it is
independent of the system parameters.

This universal feature can also be understood analytically from
multi-AR along the interface of TI and superconductor. Due to the
nature of TI the incoming spin up electron can be Andreev reflected
as spin down hole with AR amplitude $r_A$ and normal reflected as a
spin up electron with reflection amplitude $r$. Note that $r_A$ and
$r$, respectively, play the role of reflection and transmission
amplitudes for normal system. Clearly, the normal reflection
amplitude that plays the role of transmission coefficient for normal
system will vanish after $N$ successive ARs for large $N$. While for
a particular AR $r_A$ and $r$ depend on system parameters such as
$E_F$, incoming electron energy, and coupling strength $t$, it is
the multi-AR that makes the quantum Andreev effect independent of
system parameters.

\begin{figure}
\includegraphics[width=8.5cm,totalheight=6cm]{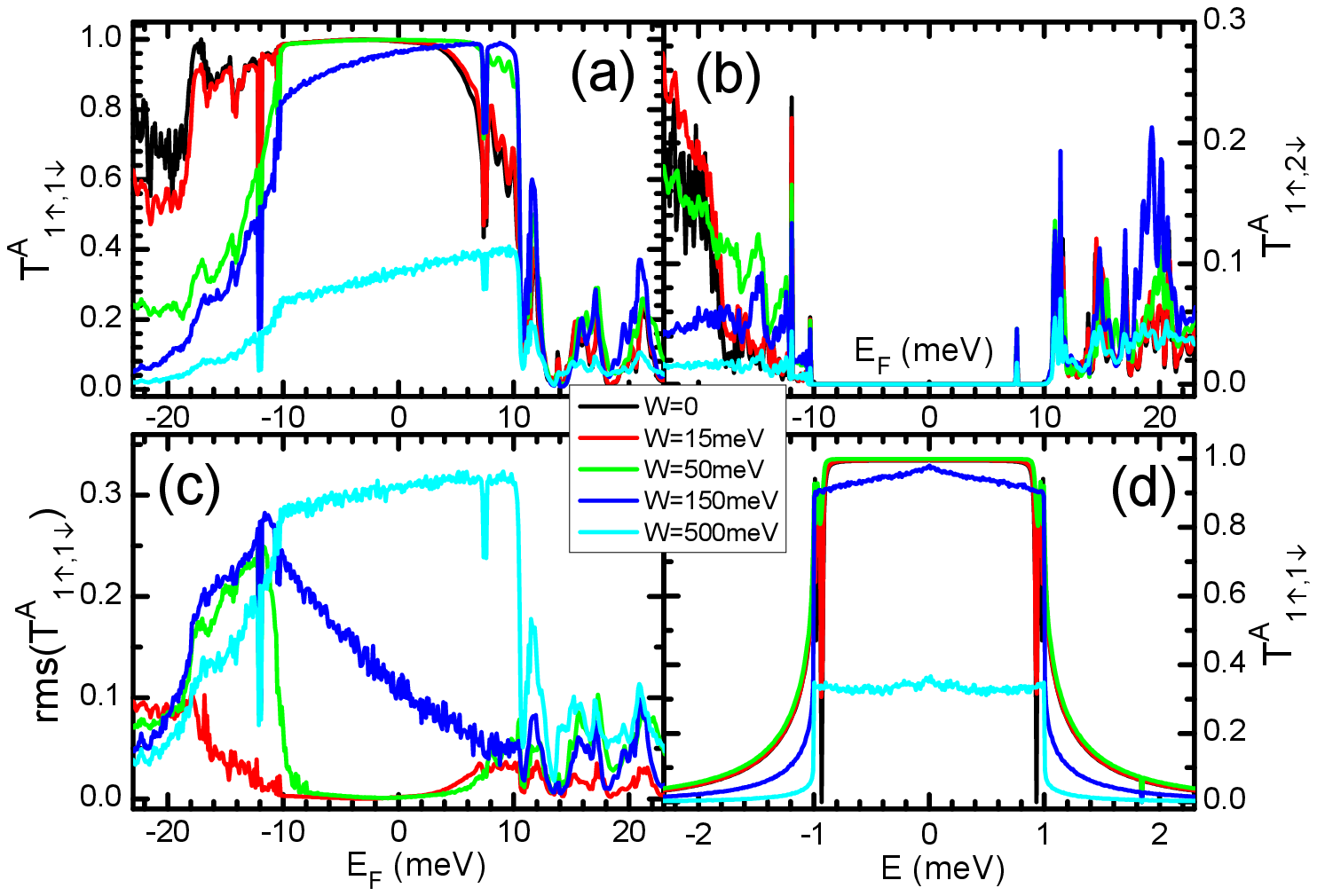}
\caption{ (Color online) (a), (b), and (c) are
$T^A_{1\uparrow,1\downarrow}$ (a), $T^A_{1\uparrow,2\downarrow}$
(b), and $rms( T^A_{1\uparrow,1\downarrow})$ (c) vs. $E_F$ with the
energy $E=0.1meV$ for different disorder strengths $W$. (d) is
$T^A_{1\uparrow,1\downarrow}$ vs. $E$ with $E_F=0$. The length of
the disorder region $L_W=150nm$ and the other parameters are the
same as Fig.1. Here all curves are averaged over up to 2000 random
configurations. } \label{fig:3}
\end{figure}

Is the quantized AR robust against the disorder? To answer this
question, we consider the on-site Anderson disorder in a region near
the TI-S interface (see the light gray (red) region in Fig.1f).
Because of the disorder, an extra on-site term $\Psi^{\dagger}_{\bf
i} \breve{w}_{\bf i} \Psi_{\bf i}$ is added on each site ${\bf i}$
in the disorder region, where $\breve{w}_{\bf i}$ is the $4\times 4$
diagonal matrix with diagonal elements ($w_{\bf i}, w_{\bf i},
-w_{\bf i}, -w_{\bf i}$). $w_{\bf i}$ is assumed uniformly
distributed in the range $[-W/2, W/2]$ with the disorder strength
$W$. Fig.3 shows the AR coefficients
$T^A_{1\uparrow,1\downarrow/2\downarrow}$ and its fluctuation versus
$E_F$ and $E$. The results show that the quantized AR plateau in
$T^A_{1\uparrow,1\downarrow}$ is very robust: the AR plateau can
persist and its fluctuation $rms(T^A_{1\uparrow,1\downarrow})$ is
zero for the disorder strength $W$ up to $100meV$ because of the
helical edge states being very robust. Upon further increasing of
$W$ from $100meV$, $T^A_{1\uparrow,1\downarrow}$ starts to decrease
and the fluctuation becomes nonzero because at large disorders the
system reaches the diffusive regime and the helical edge states are
destroyed. Hence, as long as the edge state is survived, disorder
has no effect on the quantum Andreev effect.

\begin{figure}
\includegraphics[width=8.5cm,totalheight=6cm]{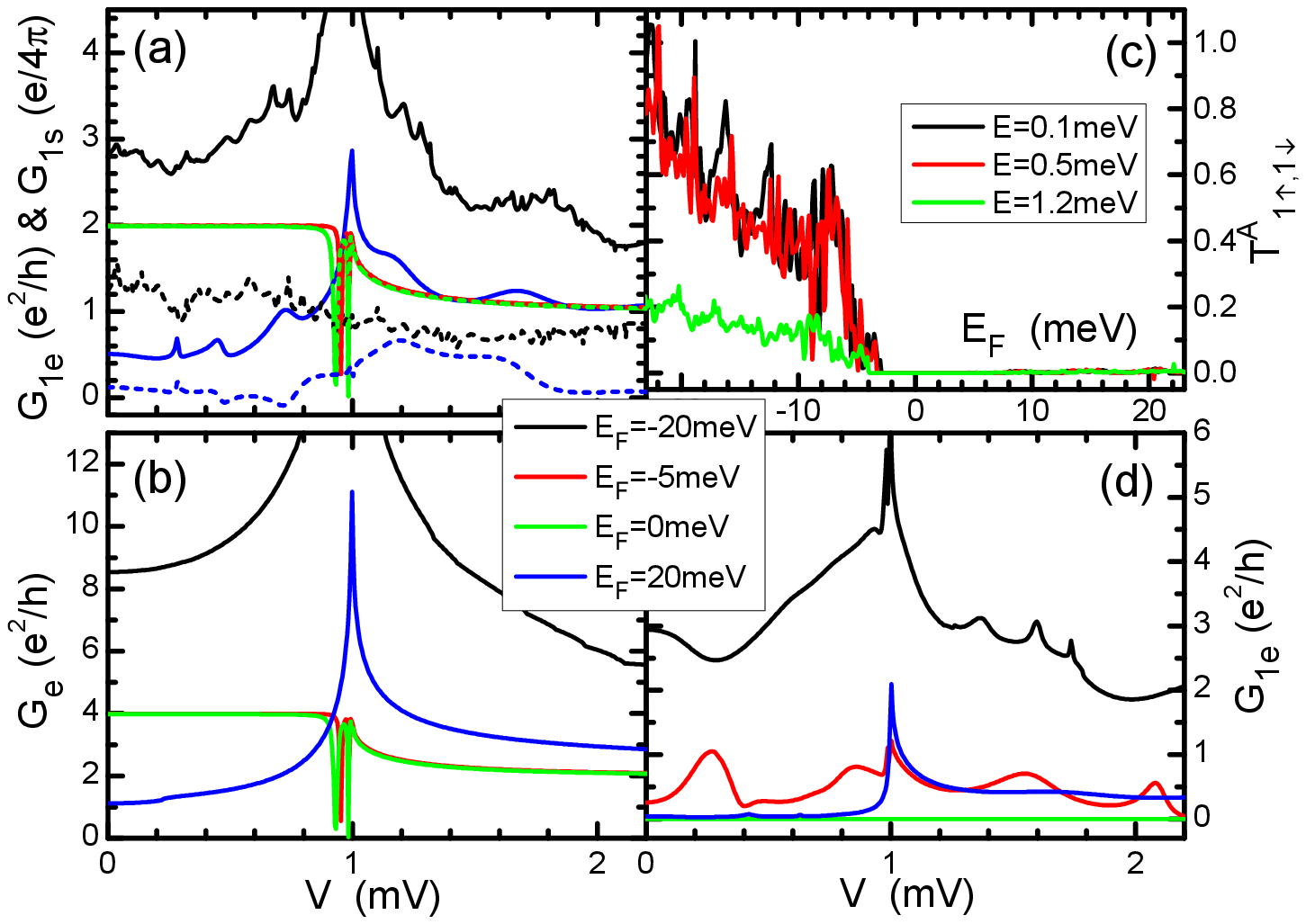}
\caption{ (Color online) (a) is the conductance $G_{1e}$ (solid
curves) and spin conductance $G_{1s}$ (dotted curves) vs. the bias
$V$ for different $E_F$ (with the legend being the same in (b)) for
the four-terminal device. (b) is the conductance $G_{e}$ vs. the
bias $V$ for the two-terminal device. (c) and (d) are
$T^A_{1\uparrow,1\downarrow}$ vs. $E_F$ (c) and conductance $G_{1e}$
vs. bias $V$ (d) for the four-terminal device with the positive
$M=2meV$. The other parameters in (a)-(d) are the same as Fig.1. }
\label{fig:4}
\end{figure}

Let us investigate the conductance $G_{ne}$ ($G_{ne}\equiv
dI_{ne}/dV$) and spin conductance $G_{ns}$ ($G_{ns}\equiv
dI_{ns}/dV$). We set the biases of the HgTe/CdTe QW terminals,
$V_1=V_2=V_3\equiv V$, and the superconductor-terminal bias $V_s=0$.
Fig.4a shows the conductance $G_{1e}$ and spin conductance $G_{1s}$
versus the bias $V$ for the four-terminal system. When $E_F$ is
inside the bulk gap, $G_{1e}$ (in the unit $e^2/h$) is exactly equal
to $G_{1s}$ (in the unit $e/4\pi$) since only the spin-up electron
traverses from the terminal-1 to the TI-S interface with the
spin-down hole Andreev reflected back. In particular, when
$V<\Delta/e$, a (spin) conductance plateau emerges with the plateau
value $2e^2/h$ ($e/2\pi$) because of the quantized AR with
$T^A_{1\uparrow,1\downarrow}=1$. Here we emphasize that like the
quantized AR this conductance plateau is also universal (i.e. it is
independent of the system parameters and the quality of TI-S
coupling) and robust against the disorder. Since the TI phase in
HgTe/CdTe QW has been realized experimentally,\cite{ref6,ref7} this
predicted quantum Andreev effect with quantized conductance plateau
should not be difficult to observe using the present technology. On
the other hand, when $E_F$ is out of the bulk gap, both $G_{1e}$ and
$G_{1s}$ are not equal and they are sensitive to the system
parameters. Finally, we comment on following two points: (i) If
HgTe/CdTe QW is in the normal state (i.e. $M>0$), no plateau emerges
for both the AR coefficient $T^A_{n\uparrow,m\downarrow}$ and
conductance $G_{ne}$ (see Fig.4c and 4d), similar to the ordinary
conductor-superconductor hybrid system. (ii). For the two-terminal
device with the HgTe/CdTe QW in TI regime, the conductance $G_{e}$
also exhibits the quantum Andreev effect with quantized AR
conductance $4e^2/h$ at $V<\Delta/e$ (see Fig.4b). However, the spin
conductance $G_{s}$ is exactly zero due to the fact that the left
and right edges of the HgTe/CdTe QW ribbon carry the same current
but opposite spin current.

In summary, we predict a quantum Andreev effect in the 2D TI-S
hybrid system, in which the AR coefficient is quantized with the
value one. Importantly, the quantized AR plateau is independent of
various system parameters and the quality of TI-S coupling, and it
also is robust against the disorders. Hence it should be easy to
observe it using the present technology. Due to the quantized AR,
the conductance also exhibits the same quantized plateau when the
bias is within the superconductor gap.

{\bf Acknowledgments:} This work was financially supported by NSFC
under Grant Nos. 10974236, 10974043, and 11074174, and a RGC Grant
(No. HKU 704308P) from the gov. of HKSAR.

\end{document}